\documentclass[reprint,amsmath,amssymb,aps,prb,showpacs,showkeys]{revtex4-1}
\usepackage{graphicx}% Include figure files
\usepackage{dcolumn}% Align table columnson decimal point
\usepackage{bm}% bold math
\usepackage{natbib}
\usepackage{color}
\usepackage{amsmath}
\usepackage{bm}            % uncommentthese two packages if you
\usepackage[T1]{fontenc}
\def \smt {Sn$_{1-x}$Mn$_x$Te}
\def \pst {Pb$_{1-x}$Sn$_x$Te}
\def \pss {Pb$_{1-x}$Sn$_x$Se}
\def \pmt {Pb$_{1-x}$Mn$_x$Te}
\def \psmt {Pb$_{1-x-y}$Sn$_x$Mn$_y$Te}

\def \pss {Pb$_{1-x}$Sn$_x$Se}
\newcommand{\angstrom}{\mbox{\normalfont\AA}}

\begin{document}
\bibliographystyle{plainnat}
% Title of the article
\title{
Band structure and 
topological phases of \psmt\ by ab initio calculations
}
% Authors
\author{ A.~{\L}usakowski}
\affiliation{Institute of Physics, Polish 
Academy of Sciences, Al. 
Lotnik\'{o}w 32/46, PL-02-668 Warszawa, Poland}
\author{P. Bogus{\l}awski}
\affiliation{Institute of Physics, Polish 
Academy of Sciences, Al. 
Lotnik\'{o}w 
32/46, PL-02-668 Warszawa, Poland}
\author{T. Story}
\affiliation{Institute of Physics, Polish 
Academy of Sciences, Al. 
Lotnik\'{o}w
32/46, 02-668 Warsaw, Poland\\
International Research Centre MagTop, Institute of 
Physics, PAS, Al. Lotnik\'ow 32/46, PL-02668,
 Warszawa, Poland}

\begin{abstract}
The change in the composition of \pst\ IV-VI semiconductor or in its 
lattice parameter can drive a transition from the topologically trivial 
to the topological crystalline insulator (TCI), crossing a region where 
the alloy is in the Weyl semimetal phase. Incorporation of the 
magnetic Mn ions induces strong perturbations of the electronic 
structure, which act on both orbital and spin variables. Our first 
principles calculations show that the presence of Mn shifts the TCI 
and the Weyl region towards higher Sn contents in \pst. When the 
Mn spin polarization is finite, the spin perturbation, like the orbital 
part, induces changes in band energies comparable to the band gap, 
which widens the Weyl area. The effect opens a possibility of driving 
transitions between various topological phases of the system by 
magnetic field or by the spontaneous Mn magnetization. We also 
propose a new method to calculate topological indices for systems 
with a finite spin polarization defined based on the concept of the 
Chern number. These valid topological characteristics enable an
identification of the three distinct topological phases of  
the \psmt\ alloy. 
\end{abstract}
\maketitle

\bibliographystyle{plainnat}
% Title of the article
\section{Introduction}

\pst\ and its selenide analogue, \pss, are IV-VI narrow gap 
semiconductors known to undergo chemical composition, pressure or 
temperature driven band inversion accompanied by the transition from 
the trivial insulator to the topological crystalline insulator (TCI)
phase.\cite{hsieh, dziawa, tanaka, xu}  
In 
contrast to (Bi,Sb)$_2$(Te,Se)$_3$ chalcogenide topological insulators, 
in TCIs it is the crystalline (mirror-plane) symmetry, not the time-
reversal symmetry, that warrants the existence of the Dirac-like states 
on specific high symmetry crystal facets of bulk rock-salt crystals.  
\cite{hsieh,ando,safaei} Experimentally, the TCI surface states were 
observed by angle-\ and spin--resolved photoemission electron 
spectroscopy \cite{dziawa,tanaka,xu,wojek}, scanning tunneling 
microscopy/spectroscopy \cite{okada}, magnetotransport and 
magnetooptical quantum oscillatory effects.\cite{dybko,wang,krizman} 
The band inversion and topological transition in \pst\  was recently 
analyzed theoretically by ourselves \cite{lusakowski2} and others 
\cite{safaei, zunger2, gao, ernst} using the density functional theory 
(DFT), the tight binding approximation (TBA), the virtual crystal 
approximation (VCA), and the method of special quasirandom structures 
(SQS) developed\cite{zunger} for the analysis of substitutional alloys. 
Confirming the successful basic picture obtained in early VCA 
calculations \cite{safaei} the other methods accounted for local 
chemical disorder, inevitably present in alloys. It was discovered that 
due to the splitting of the bands induced by locally varying crystal 
field 
symmetry there exists a transition region between the trivial and the TCI 
phase characterized by the zero energy gap \cite{lusakowski2} with a 
possible Weyl semimetal type of energy bands 
arrangement.\cite{zunger2} 

In the quest for efficient ways of controlling  the topological 
insulators 
and semimetals their alloying with other semiconductor materials 
proved very promising. \cite{ando}  Particularly interesting is alloying 
of 
TCIs with magnetic semiconductors, like MnTe or MnSe, thus combing 
the topological and magnetic properties of materials. Recent 
developments in the field of topological materials  showed a rapid 
progress in ferromagnetic (FM) and antiferromagnetic (AFM) 
heterostructures expected to exhibit quantum anomalous Hall effect 
and topological magneto-electric effect.\cite{ando,qahe,natur} There 
exist several proposals for controlling topology in TCIs with nonzero 
magnetization as well for topological transition from the TCI to the Weyl 
semimetal (WSM) state.\cite{liu, serbyn, reja, bonnani}
\\
\pst\ with Mn is known as IV-VI diluted magnetic (semimagnetic) 
semiconductor (DMS) exhibiting carrier-induced ferromagnetism driven 
by the Rudermann-Kittel-Kasuya-Yosida (RKKY) indirect exchange 
interaction via holes. \cite{story11086,eggenkamp} As the  solubility 
limit of Mn in bulk crystals of \psmt\ is about 12 at. \%, the 
ferromagnetic transition temperature observed in the bulk crystals is 
below 30~K.\cite{eggenkamp,lazar,mazur} For thin epitaxial layers of 
topologically nontrivial terminal alloy, \smt, the solubility appears 
smaller due to lower temperatures required for epitaxial growth, and 
the ferromagnetic Curie temperature is below 10~K. 
\cite{nadolny,bonnani} Importantly, Mn in \pst\ substitutes Sn$^{2+}$ 
or Pb$^{2+}$ ions as isoelectronic Mn$^{2+}$ ion with the configuration 
$3d^5$ and magnetic moment of 5 Bohr magnetons, as verified by 
electron paramagnetic resonance  studies of both very diluted 
paramagnetic crystals \cite{story1996} and more concentrated 
ferromagnetic ones. \cite{epr-story1993} 
\\
The influence of Mn ions on the band structure of quaternary 
system \psmt\ and corresponding terminal ternary alloys \pmt\ 
and \smt\ was studied to explain their very good thermoelectric 
parameters as well as ferromagnetic properties. It was found,  both 
theoretically and experimentally, that the key role is played by  the 
band 
of heavy holes ($\Sigma$-band) and the position of its top with respect 
to the top of the highest valence band located at the L-point of the Brillouin 
zone.\cite{lusakowski1,bukala,tan,tan1,he,tian,znalezc}  A good 
agreement was achieved between theoretical predictions and optical 
and thermoelectric data for the topologically trivial \pmt\ alloy: with 
the increasing Mn content the main gap at the L-point increases while  the 
energy separation between the L- and the $\Sigma$-bands 
decreases.\cite{lusakowski1,bukala,tan} Theoretical studies of the band 
structure of the TCI \smt\ agreed on decreasing energy separation of 
the L- and $\Sigma$-bands, but provided conflicting predictions for main 
gap at the L-point, finding either its increase\cite{tan1,he,tian}  or 
decrease\cite{znalezc} with the increasing Mn content. The band 
structure of \psmt\ in the band inversion region was not studied 
theoretically with ab initio methods. In the early model analysis of 
temperature, composition and carrier concentration dependence of 
thermoelectric power of \psmt\ it was assumed that the main gap at the 
L-point increases with the increasing Mn content,\cite{lazar1}  as 
typically observed in known II-VI and IV-VI DMS materials. 
\\
In the present work, we employ first principles calculations to study the 
influence of the Mn ions on the electronic structure of \psmt. 
Incorporation of magnetic Mn ions into such crystals, apart from the changes in 
chemical composition and local crystal symmetry, leads in 
general to breaking the time reversal symmetry. A finite spin 
polarization of the Mn sublattice induces large spin splittings of the 
band states, which can be comparable to the band gap energy. 
 
By analyzing the band structure, topological indices and charges, as well 
as the Weyl's nodes we demonstrate how changes in the chemical 
composition, or in the lattice parameter (by pressure or strain) induce a 
transition from the trivial to the TCI phase of \psmt. Typically, the 
transition passes through an intermediate region characterized by zero
energy gap, corresponding to the Weyl semimetal  
phase. Importantly, we predict that the transition from the trivial to 
the 
Weyl phase can be driven by the magnetic field or by spontaneous 
magnetization. 

The response of a band extremum to the perturbation induced by Mn 
depends on its symmetry, and it is different for the states
derived from the L$_6^+$ and the L$_6^-$ band extrema of PbTe and SnTe. 
This holds for both the spin and the orbital perturbation. 
We trace this effect 
to the symmetry-dependent hybridization between the $p$(Te) and the 
$s$(Mn) and $d$(Mn) orbitals. 

The presence of the Mn ions leads to the increase of the band gap in the 
trivial region, but to the decrease of the absolute value of the inverted 
gap in the  TCI phase. This result is in agreement with experimental 
observations available for the trivial phase, e.g., in  Pb$_{1-
x}$Mn$_x$Te.\cite{bpz} The calculated impact of Mn on the inverted 
gap in the TCI phase is particularly relevant. Indeed, in this case 
analysis 
of experimental data (see, e.g., Ref. \onlinecite{lazar1}) is obscured by 
contradictory theoretical predictions,\cite{tan1,he,tian} and the 
conclusions depend on the specific band structure models adopted in 
the interpretation.

\section{Technical details of calculations} 
\subsection{Modeling of disordered
  P\MakeLowercase{b}$_{1-x-
y}$S\MakeLowercase{n}$_{x}$M\MakeLowercase{n}$_{y}$T\MakeLowercase{e}
  crystals}

Infinite 
P\MakeLowercase{b}$_{1-x-
y}$S\MakeLowercase{n}$_{x}$M\MakeLowercase{n}$_{y}$T\MakeLowercase{e}
random mixed crystals are modelled by the $2\times 2\times 2$ supercells 
containing 64 atoms: 32 Te anions and 32 Pb, Sn or Mn cations. 
In the following, 
instead of \psmt\ with the specified $x$ and $y$ we often use  the 
notation 
Pb$_{k}$Sn$_{l}$Mn$_m$Te$_{32}$ where $k$, $l$ and  $m$, 
$(k+l+m=32)$, are the numbers of Pb, Sn and Mn atoms  in the 64 
atoms supercell, respectively.
The most important problem is the choice of the spatial distribution of 
cations in the supercell, because, as we showed in Ref. 
\onlinecite{lusakowski2} for \pst, the band structure and in particular 
the energy gap $E_{gap}$ strongly depend on the cation 
configuration. 

To solve this problem, we applied  the 
SQS approach.\cite{zunger} The aim of this approach is to find 
such positions of different cations in the supercell that their 
distribution resembles a random distribution of cations in the infinite 
alloy as much as possible for several coordination spheres. 
However, for three 
different cations in the supercell, Pb, Sn and Mn, the number of their 
possible configurations is much larger than in the case of Pb and Sn 
only, thus the probability of finding the best possible SQSs is much 
smaller. In spite of this, as it will be seen in the following, although 
the calculated dependencies of energy gaps on the Sn 
concentrations are not perfectly smooth, general trends can be easily 
resolved.
\\
For a given distribution of cations, the starting point of our analysis 
are the DFT calculations performed with the  
open-source OpenMX package.\cite{openmx} The calculations were 
done using the local density approximations with the 
Ceperly-Alder\cite{CA} exchange-correlation functional. For Sn and Mn we 
used pseudopotentials distributed with the OpenMX (version 
2013), for Pb and Te we used pseudopotentials with 4 and 6 valence 
electrons, respectively, generated previously using the program 
ADPACK distributed with the OpenMX. All the input 
parameters for calculations of pseudopotentials for Pb and Te were 
described in Ref.~\onlinecite{lusakowski1}.
\\
In the calculations we assume the experimental dependence of 
the lattice parameter on the chemical composition:\cite{miotkowska}  
\begin{equation}
  \label{sm}
  a=6.460-0.145x - 0.558y\ \ \angstrom.
\end{equation}
The experimental equilibrium Mn-Te bond length, $d_{Mn-Te}\approx 
2.96$~\AA, is much 
smaller than $d_{Pb-Te}\approx 3.23$~\AA\ and $d_{Sn-Te}\approx 
3.16$~\AA.\cite{iwanowski} 
Consequently, the ions in the alloy do not occupy perfect NaCl lattice 
sites, and the internal  
distortions are of importance when considering the alloy band gap. We 
will show below that 
the hybridization between $3d$(Mn), $4s$(Mn) and $5p$(Te) orbitals has 
important influence 
on the energy gap. This effect depends on the distance between Mn and Te, 
making the 
geometry optimization necessary. 
\\
The energy gaps presented in the figures below are the minimal direct 
energy gaps on the 
[111] direction in the 3D Brillouin zone (BZ). However, one should keep 
in mind that, contrary 
to PbTe or SnTe, due to lack of $O_h$ local symmetry in most of the 
considered systems the 
smallest energy gaps are  not, in general, placed on this direction (see 
the next section for an example). 
\\ 
The calculations are mostly done for systems containing even numbers of 
Mn ions in the supercell. With this choice one can study the impact of Mn 
doping on \pst\ in the paramagnetic case, by assuming AFM spin 
configurations with the vanishing total spin, and separate the effects of 
spin polarization, modelled by assuming the FM spin configurations. In 
the 
latter case, the valence and conduction bands are spin-split, which in 
turn 
significantly affects the energy gaps, particularly when they are very 
small. 
Both situations are accessible experimentally, since the Curie 
temperature 
in the $p$-doped tin tellurides is of the order of 10 K.

\subsection{Calculations of topological indices} 

The OpenMX package enables to obtain TBA parameters for the TBA 
Hamiltonian. Using these parameters 
we calculate the necessary  topological indices: the Chern numbers 
(CNs), the spin Chern numbers (SCNs), and the numbers  
$C_{s+}$ and $C_{s-}$.\cite{prodan} The method of calculation of
$C_{s+}$ and $C_{s-}$ was described in
Ref. \onlinecite{lusakowski2}. In short, the idea is to divide the
valence band states for every ${\bm k}$ in the BZ into two sets,
$P_+({\bm k})$ and $P_-({\bm k})$. In $P_+({\bm k})$ ($P_-({\bm k})$) 
there are states
with positive (negative) average spin. Taking the sums over ${\bm k}$ we 
obtain two vector bundles $P_+=\oplus_{\bm k} P_+({\bm k})$
$P_-=\oplus_{\bm k} P_-({\bm k})$. Calculating the Chern numbers for
these bundles we obtain $C_{s+}$ and $C_{s-}$. The spin Chern number
$SCN=(C_{s+}-C_{s-})/2$. 
The  method of calculation of $C_{s\pm}$ is based on the approach 
proposed 
by Fukui {\it et al.}  \cite{suzuki}
\\
The microscopic Hamiltonian, i.~e., the Pauli--Schr\"odinger
Hamiltonian for electrons in \psmt\ is   
time reversal  invariant. However, after the DFT
calculations the time reversal invariance is broken. The resulting
spins of ions  
are finite and consequently the resulting TBA Hamiltonian's  have no time 
reversal 
symmetry. This feature was directly checked in a number of cases. 
Therefore, the general theorem that the Chern number calculated for a 
given 2D plane in the 3D BZ should be zero is not valid. 

The BZ of our supercells is a cube. In the 
calculations of topological indices we use the (001) plane crossing the 
${\bm k}$=(0,0,0) $\Gamma$  point. This point corresponds to the $L$ 
points of the unfolded BZ, and the main energy gap is situated at 
$\Gamma$ or in its close vicinity. 
 
The procedure for calculation of SCNs described 
previously\cite{lusakowski2} can be applied directly to the systems 
without magnetic ions, or those containing an even number of  magnetic 
ions in the supercell, arranged in such a way that their total magnetic 
moment vanishes.  
Then, the dimensions $n_{\pm}({\bm k})$ of $P_{\pm}({\bm k})$  are equal,
$n_+({\bm k})=n_-({\bm k})$. 
In the case when the total spin of the system is nonzero the 
procedure must be modified. 
In general, $n_+({\bm k})\equiv n_+$
and $n_+({\bm k})\equiv n_-$
are independent on ${\bm k}$, 
and they are related by  
$n_+({\bm k})=n_-({\bm k})+N_S$ where $N_S=5(N_{Mn\uparrow}-
N_{Mn\downarrow})$; exceptions are discussed below.
Here $N_{Mn\uparrow}$ and $N_{Mn\downarrow}$ are the 
numbers of Mn atoms in the supercell with the spin up and down, 
respectively, and the factor 5 is related to the number of 
unpaired spins on the $3d$(Mn) shell. Thus, 
\begin{equation}
  \label{eq1}
n_-=(n_{val}-N_S)/2\ \ \ \ \ \ 
n_+=(n_{val}+N_S)/2 , 
\end{equation}
where $n_{val}$ is the number of occupied states. 
This is related to the exchange polarization of the bands due to 
nonvanishing net magnetization caused by Mn ions. Next, we 
divide the valence states into two subspaces and calculate Chern 
numbers $C_{s+}$ and $C_{s-}$. 
\\ 
This procedure was applied successfully in a vast majority of 
cases, however, for completeness, one should mention possible 
problems. 
Sometimes, the calculated $C_{s+}$ and $C_{s-}$ are 
noninteger.  For example, for  Pb$_{31-
n}$Sn$_{n}$Mn$_{1}$Te$_{32}$ meaningful results 
(integer $C_{s\pm}$) are obtained only for trivial and nontrivial 
regions, $n 
\le 9$ and $n \ge 17$, respectively. In the transition region $10\le n 
\le 16$, 
where the values of the energy gaps are nearly  zero, the procedure 
fails. This is related to the fact that in such  cases there are a few 
points in the 2D plane of the BZ where the number of states with 
positive and negative average spins are not equal to $n_+$ and 
$n_-$ as in Eq. \ref{eq1} but equal to $n_++1$ and 
$n_--1$, respectively. Thus it is impossible  
to build  two vector bundles $P_+$ and $P_-$ of the dimensions $n_+$ and 
$n_-$, respectively and to  calculate corresponding $C_{s+}$ and 
$C_{s-}$.  The reason of those problems is the practically vanishing 
$E_{gap}$. A detailed analysis shows that in the transition 
region the system is in the Weyl semimetal phase, and sometimes the 
Weyl's nodes are at the plane in ${\bm k}$-space used to 
calculate $C_{s+}$ and $C_{s-}$. Of course, for systems with 
$E_{gap}$=0 the calculations of CNs do note make sense. 
The example of Pb$_{15}$Sn$_{16}$Mn$_{1}$Te$_{32}$ is 
discussed in the next Section. 

\subsection{Calculations for Weyl semimetal region}
The analysis of Weyl semimetal phase is not an easy numerical task
because the zero  
energy gap points are grouped in the close vicinity of the $\Gamma$ 
point in the supercell BZ. In the studied cases, all the Weyl nodes are 
contained 
in a cube of dimension 0.02~\AA$^{-1}$. 
In few cases we identified the points in the BZ where $E_{gap}=0$. 
However, in most cases we applied a much faster although less accurate 
method. Namely, the cube with the edge of 0.02~\AA$^{-1}$ was 
divided into 1000 smaller cubes with edges 0.002~\AA$^{-1}$. Next, 
for all small cubes we calculated the Berry flux through their faces. Of 
course, in this procedure we miss the cases where a pair of Weyl nodes of 
opposite charges is present inside a small cube.

\section{\label{results}Results and discussion}

\begin{figure}
\includegraphics[width=\linewidth]{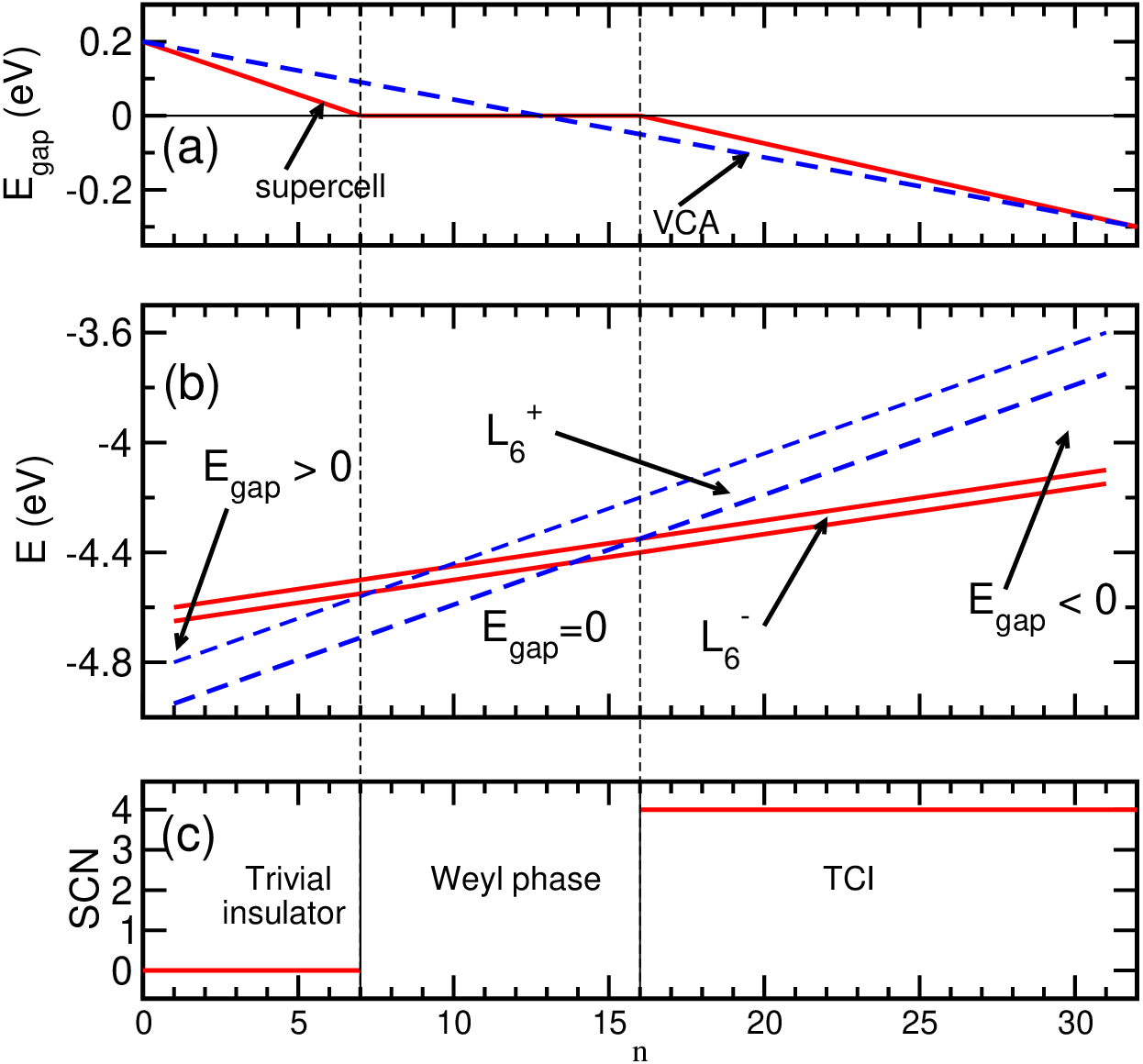}
\caption{\label{fig1} 
(color online) Schematically shown composition dependence of the relevant 
physical parameters of the \pst\ alloy. n is the number of Sn atoms in 
the   Pb$_{32-n}$Sn$_{n}$Te$_{32}$ supercell. (a) Energy gap 
calculated within the supercell method:  continuous line, and 
with the VCA: broken line. (b) Splittings of the energy 
levels at the $\Gamma$ point for the supercell BZ.  The energies of  
$L_6^-$-- and the $L_6^+$--derived states are between continuous red 
and broken blue lines, respectively. 
(c) Spin Chern number for the trivial 
and the TCI phase separated by the Weyl semimetal phase. } 
\end{figure}

We begin with a brief summary of the main features characterizing the 
electronic structure of \pst\cite{lusakowski2} without Mn ions. They are 
presented in Fig. \ref{fig1}, which schematically shows the composition 
dependence of the band gap together with the relevant energies of the 
valence and conduction bands at the $\Gamma$ point of the supercell BZ 
and the topological indices. 
\\
With the increasing Sn content, $E_{gap}$ changes the character from 
positive in PbTe to negative in SnTe, Fig.~\ref{fig1}a, driving the 
system 
from topologically trivial to TCI, which is also reflected in the non-
vanishing 
spin Chern numbers in the Sn-rich \pst. Qualitatively different 
characters of 
the transition are obtained in the VCA and in the supercell method used 
in 
this paper. 

In the VCA, the chemical disorder of an alloy is absent, the system 
retains 
both the $O_h$ point symmetry and the translational symmetry of the rock 
salt structure. As a result, the transition between topologically trivial 
and 
nontrivial phases is sharp, and takes place at a well-defined critical 
composition. In the supercell method, the alloy is simulated by the 
repeated supercells. The L points of the PbTe (or SnTe) Brillouin zone 
are 
folded to the $\Gamma$ point of our 64-atom supercell BZ. After the 
folding, both the L$_6^+$- and the L$_6^-$-derived bands of pure PbTe are 
8-fold degenerate. These degeneracies are lifted in the presence of two 
types of cations because of the different chemical nature of Pb and Sn, 
and 
of the disorder in their spatial distribution. This splitting is 
schematically 
shown in Fig.~\ref{fig1}b. The magnitude of splittings and possible final 
degeneracies (e.g., double degeneracies in systems with inversion 
symmetry discussed below) depend on the actual distribution of Pb and Sn 
in the supercell. Consequently, in the supercell approach the trivial-TCI 
transition is smeared, there is a relatively wide composition window in 
which the band gap between occupied and unoccupied states vanishes, and 
the system is in the WSM phase, \cite{zunger2} where 
topological indices are in general not defined. 
\\
The energy of the $L_6^-$ relative to the $L_6^+$ band extremum can also 
be reduced by application of the hydrostatic pressure. This closes the 
positive gap of PbTe, opens the negative gap of SnTe, and can drive the 
pressure-induced transition from the trivial to the TCI phase in \pst. 
Again, 
the splitting of the energy levels leads to a smeared character of the 
transition, which proceeds through the Weyl phase (see Ref. 
\onlinecite{zunger2} for details). In the VCA the transition is 
sharp, and the WSM phase is absent. \\
Remarkably, the existence of semimetal region in \pst\ was
proposed based on recent experimental studies of temperature and
composition dependence of conductivity.\cite{zhang}

\subsection{\label{results1}Comparison of
P\MakeLowercase{b}$_{30}$S\MakeLowercase{n}$_{2}$T\MakeLowercase{e}$_{32}
$  and 
P\MakeLowercase{b}$_{30}$M\MakeLowercase{n}$_{2}$T\MakeLowercase{e}$_{32}
$}

Mn ions in \pst\ are magnetic and assume the high spin state S=5/2. The 
substitution of Mn for a cation introduces a perturbation acting on both 
the orbital and the spin variables, which in the following are referred 
to 
as the chemical and the spin part of perturbation, respectively. Both 
effects are analyzed here. The spin perturbation affects the band 
structure only when the Mn sublattice is spin polarized. Accordingly, we 
will assume a finite spin polarization of Mn (present in the FM phase, or 
induced by a weak magnetic field), but its influence on the orbital 
motion (including the Landau quantization) is neglected, because it 
requires a different approach to the band structure calculations, such as 
the effective mass model. 
%\onecolumngrid

\begin{figure*}
 \includegraphics[width=0.7\linewidth]{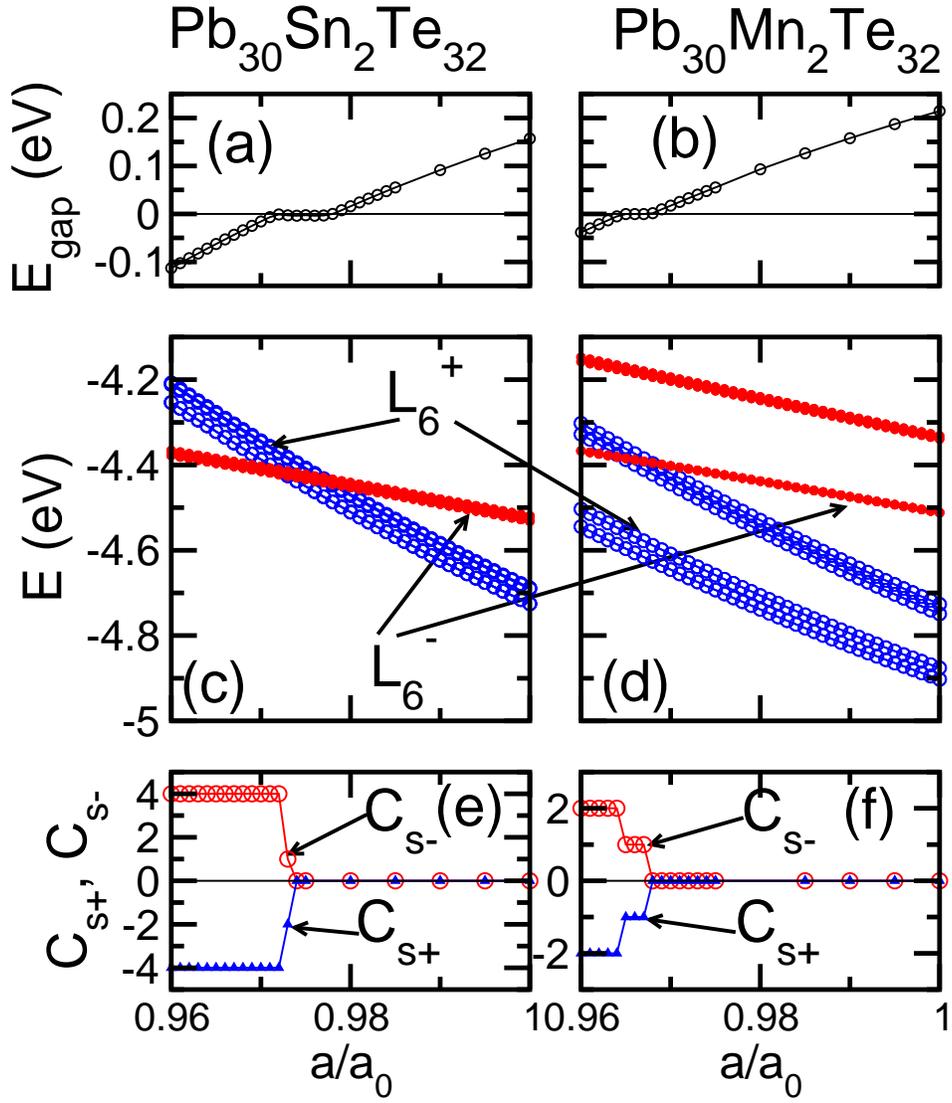}
\caption{\label{fig2} 
(color online) Dependence on the lattice parameter $a$ of: 
(a), (b)  energy gap,  (c), (d) splitting of the $L_6^+$ and $L_6^-$ 
derived levels 
at the $\Gamma$ point of the supercell BZ, 
and (e), (f) the Chern numbers $C_{s+}$, $C_{s-}$.
The left panel corresponds to Pb$_{30}$Sn$_{2}$Te$_{32}$, the right one 
to Pb$_{30}$Mn$_{2}$Te$_{32}$. 
$a_0$ is the equilibrium lattice parameter of PbTe.}
\end{figure*}
%\twocolumngrid

We first neglect the spin perturbation, and compare the impact of 
doping PbTe with Sn and Mn. The comparison reveals the main features 
also present in the remaining cases. The results, obtained for a 64-atom 
PbTe supercell containing two Sn ions and that with two Mn ions with 
antiparallel spins, are presented in Fig.~\ref{fig2}. The Figure shows 
the 
effect of hydrostatic pressure (monitored by the decrease of $a$) on the 
electronic structure. The positions of Sn and Mn atoms in the supercells 
are  the same, and the internal relaxations are neglected. 
\\
From Figs.~\ref{fig2}a and ~\ref{fig2}b it follows that the incorporation 
of Mn ions to PbTe at equilibrium $a=a_0$=6.46~\AA\ leads to the 
increase of the band gap from 0.20 eV for PbTe to 0.214~eV for 
Pb$_{30}$Mn$_{2}$Te$_{32}$, while the incorporation of Sn decreases 
$E_{gap}$ to 0.16 eV in Pb$_{30}$Sn$_{2}$Te$_{32}$ in accord with 
measurements.\cite{nimtz, khoklov} Considering the pressure 
dependence of the band gap we find that for both 
Pb$_{30}$Sn$_2$Te$_{32}$ and Pb$_{30}$Mn$_2$Te$_{32}$, 
$E_{gap}$ decreases with the decreasing lattice constant $a$ (i.e., with 
the increasing hydrostatic pressure), and eventually changes sign to 
negative, again in agreement with experiment.\cite{bpz, nimtz, khoklov, 
justyna} 
\\
The pressure dependence of $E_{gap}$ shows regions where the band 
gap practically vanishes, and both systems are in the WSM 
phase. This effect was already pointed above, and is 
related to the alloy broadening, i.e., to the splitting of energy bands 
in 
mixed crystals. The splitting is shown in some detail in Figs. 
\ref{fig2}c 
and \ref{fig2}d. At the $\Gamma$ point of our 64-atom supercell BZ, the 
L$_6^+$- as well as the L$_6^-$-derived bands of PbTe are 8-fold 
degenerate. These degeneracies are lifted in the presence of the Sn or 
Mn ions. Comparing the results for Pb$_{30}$Sn$_{2}$Te$_{32}$ and 
Pb$_{30}$Mn$_{2}$Te$_{32}$ we see that the level splittings, and thus 
the alloy broadening of the energy spectrum, is much larger in the latter 
case. This is because the substitution of Pb by a Group-II Mn ion 
generates a stronger crystal and electronic perturbation then the 
substitution of Pb by the isoelectronic Sn. 
\\
Finally, Figs. \ref{fig2}e and \ref{fig2}f show the topological indices 
for 
the two alloys. The sign of the energy gaps in Figs. \ref{fig2}a and 
\ref{fig2}b is determined by zero or nonzero values of the SCN. Again, in 
agreement with Figs. \ref{fig2}c and \ref{fig2}d, the Weyl region 
between the TCI and the trivial phases is broader in 
Pb$_{30}$Mn$_{2}$Te$_{32}$, although it is less clear in 
Figs. \ref{fig2}a and \ref{fig2}b. 
\\
\subsection{\label{ss2}Influence of Mn on the band structure of 
  Pb$_{30}$Mn$_{2}$Te$_{32}$ and Sn$_{30}$Mn$_{2}$Te$_{32}$.}

\begin{figure}
  \includegraphics[width=\linewidth]{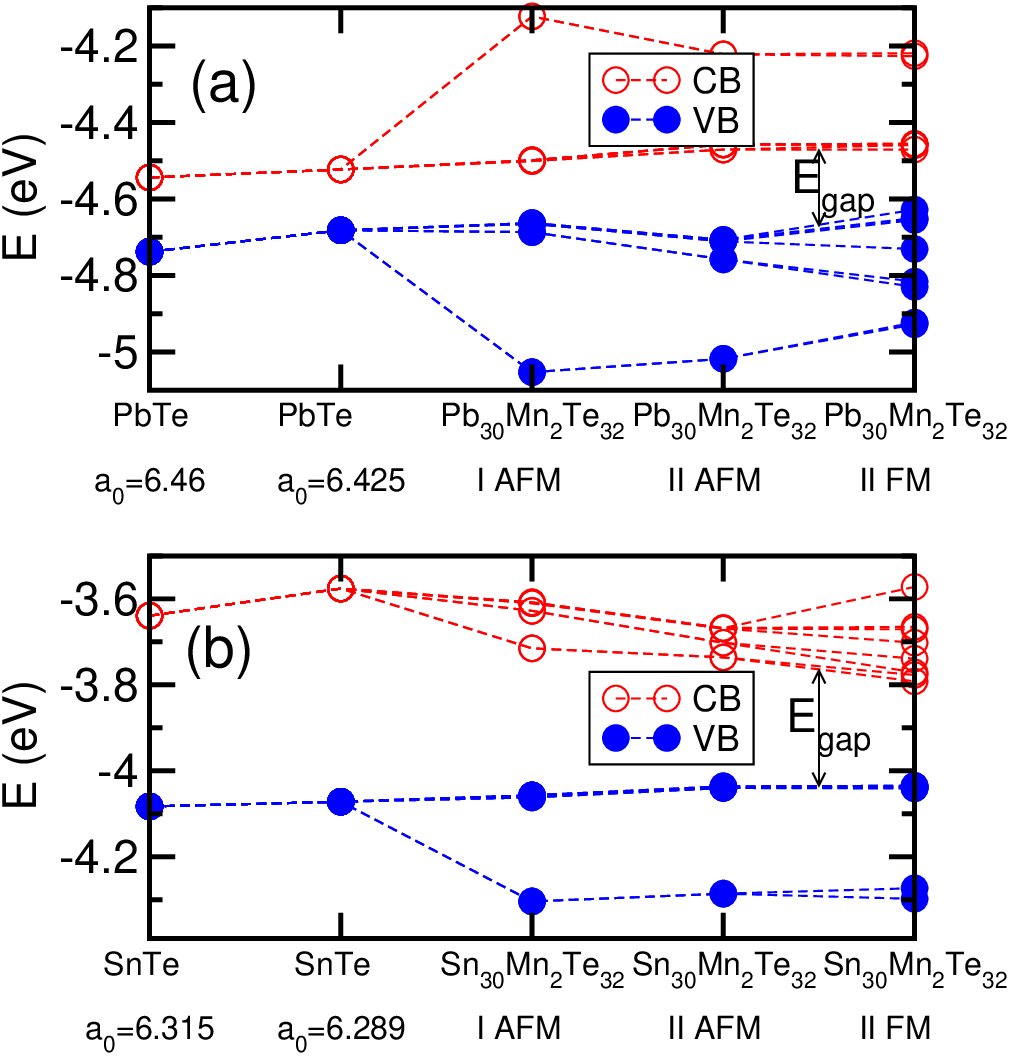}
  \caption{\label{fig3} (color online) 
The energy levels of the highest valence and the lowest conduction bands 
at the $\Gamma$ point of the supercell BZ for (a) PbTe and (b) SnTe with 
2 Mn ions. 
The consecutive steps correspond to the adjustment of the lattice 
parameter, introduction of Mn in the AFM arrangement without (I) and with 
(II) the lattice relaxation, and to the FM spin arrangement.} 
\end{figure}

The influence of Mn ions on the band structure of PbTe and SnTe is 
analyzed in Fig. \ref{fig3}, which shows the energy position of  8
highest valence bands and  
the 8 lowest conduction bands. As before, we use supercells 
with the lattice parameters taken according to Eq. \ref{sm}, and 
containing two Mn ions. The Mn ions are placed at (0, 0, 0) and $a_0$(1, 
1, 1) in a perfect rock salt lattice neglecting the internal distortions. 
Thus, the Mn ions form a cubic body centered lattice, and the system 
has the $O_h$ symmetry. With this choice the wave functions are either 
even or odd with respect to inversion (like the wave functions of the 
L$_6$ band extrema in PbTe and SnTe), which makes the analysis more 
transparent. We mention that the calculated band gaps differ from 
those obtained with the SQS method, which reflects the dependence of 
$E_{gap}$ on the Mn distribution.  The choice of supercells with an even 
number of Mn ions allows for separation of the effects induced by the 
chemical and the spin perturbation, which is achieved by comparing the 
FM and AFM spin configurations. 
\\
The final band structure of both systems is achieved in 4 steps. First, 
the 
pure compound is dilated or compressed to the appropriate lattice 
constant. In the second step, Mn ions are introduced to the supercell in 
the AFM spin configuration, but the atoms are not allowed to relax, 
which is denoted as the configuration I. We see that for both systems 
the Mn chemical perturbation induces splittings of the 8-fold 
degenerate band extrema by 0.1-0.3 eV, which is comparable to 
$E_{gap}$ of the PbTe and SnTe hosts. 
\\
In the third step, atoms relax to the configuration II, in which the 
nearest Te neighbors are moved towards the Mn ions along the bond 
directions, so that the Mn-Te bonds are reduced from 3.23 to 2.95~\AA. 
The latter value is equal to that obtained after the geometry 
optimization and, in the case of \smt\, is very close to the experimental 
one.\cite{iwanowski} This effect leads to small changes in band 
energies. In the last step we assume the FM spin arrangement. Inclusion 
of the spin polarization reduces $E_{gap}$. In both PbTe and SnTe, the 
spin splittings are large, showing that the spin and the chemical 
perturbation are equally important, and can be comparable to the band 
gap. 
\\
Interestingly, as it follows from Fig. \ref{fig3}, the response to the Mn 
perturbation depends on the band symmetry. In particular, the spin 
splittings are considerably more pronounced in the case of the bands 
derived from $L_6^+$ than those derived from the $L_6^-$ band 
extrema, independent of the host and of the valence or conduction 
band character. We relate this result with the atomic orbitals composition of the 
corresponding wave functions. In the case of 
Pb$_{30}$Mn$_{2}$Te$_{32}$, the wave function of the VBM 
are even with respect to the inversion symmetry 
operation. They contain contributions from the $3d$(Mn) orbitals, and 
these contributions are larger for the configuration II then for the 
configuration I. This is due to the stronger hybridization of $3d$(Mn) 
with $p$(Te) after the lattice relaxation, when the Mn-Te bonds are 
shorter. The wave functions of the $L_6^-$-derived conduction levels 
are odd with respect to the inversion, and practically do not contain the 
Mn orbitals and do not respond to the Mn presence. The situation is 
different in Sn$_{30}$Mn$_{2}$Te$_{32}$. The wave function of the 
VBM are odd and do not contain the Mn orbitals. The wave function of 
the lowest valence level shown in the Fig. \ref{fig3} is even with respect to inversion and contains the 
4$s$(Mn) orbital. The wave functions of the conduction states are even. 
The first conduction level contains $4s$(Mn), while the higher three 
levels contain the $3d$(Mn) orbitals. \\

\subsection{\label{results2}Supercells with 0, 2 or 4 
manganese ions}

In this Section we analyze the dependence of the energy gap and 
topological properties of \psmt\ on the Sn concentration. The impact of 
Mn is revealed by comparing three cases: that with no Mn, $y=0$, with 
$y=0.0625$, and with $y=0.125$, which corresponds to 0, 2, and 4 Mn 
atoms at the cation sites of a 64-atom supercell, respectively. We begin 
with the antiferromagnetic configuration of Mn spins.
\begin{figure}
  \includegraphics[width=\linewidth]{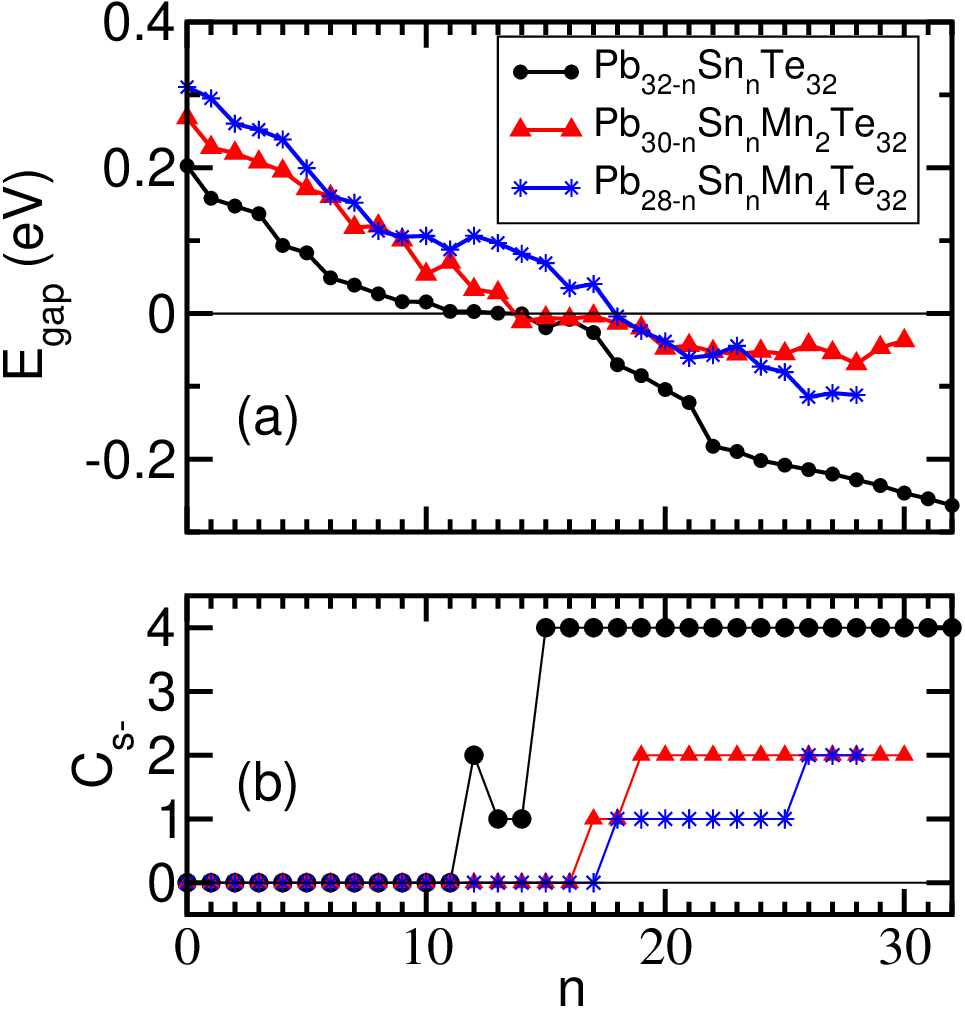}
  \caption{\label{fig4} (color online) Dependence on the number of 
Sn ions in the
    supercell of: (a) the energy gaps, (b) the spin 
Chern number
    $C_{s-}$  for 
    Pb$_{32-n}$Sn$_{n}$Te$_{32}$ (dots), for
    Pb$_{30-n}$Sn$_{n}$Mn$_{2}$Te$_{32}$ (triangles) 
and Pb$_{28-n}$Sn$_{n}$Mn$_{4}$Te$_{32}$ (stars).}
\end{figure}
In Fig. \ref{fig4} we show $E_{gap}$ and the corresponding topological 
indices as the function 
of the Sn concentration. The increase of $E_{gap}$ with the increasing Mn 
content in PbTe 
was analyzed above. As we pointed out, doping PbTe with Mn lowers the 
energy of the 
$L_6^+$ band relative to the $L_6^-$ band, thus increasing the band gap. 
In SnTe this effect 
takes place as well, but in this case it decreases the inverted 
$E_{gap}$. As it follows from Fig. 
\ref{fig4}, this effect of Mn persists also in the \psmt\ alloy in the 
whole composition range.  
\\
While the decrease of energy of $L_6^+$ relative to that of $L_6^-$ 
induced by 2 and 4 Mn 
ions in the supercells is clear for all Sn concentrations, it is not 
always possible to distinguish 
the results for 2 and 4 Mn ions. This problem stems from the fact that 
the differences in band 
energies, and in particular the band gap itself, are smaller than the 
fluctuations inherent to 
our approach. Indeed, they could be got rid of by averaging over a 
substantially larger number 
of atomic configurations in the supercells. Apparently, in this specific 
case, the SQS 
approximation is not accurate enough. 
\\
In spite of that problem, two qualitative effects should be noticed. 
First, we find that as a 
result of the Mn-induced change of $E_{gap}$ in \psmt\ the trivial-to-
nontrivial transition 
region is shifted to higher concentrations of Sn. Our calculations 
indicate that for \pst\ the 
energy gap vanishes for $x\approx 0.35$, what corresponds quite well to 
the experimental 
values $x\approx 0.4$. For \psmt\ with $y\approx 0.06$, $E_{gap}=0$ for a 
higher $x\approx 
0.5$, but the precise experimental value is not known. 
\\
Second, since adding Mn enhances alloy broadening, one would expect that 
the region of the  
Weyl phase is wider in \psmt\ then in \pst. Paradoxically, the calculated 
$E_{gap}(x)$ 
dependence exhibits an opposite effect, and in the case of 4 Mn ions in 
the supercell the Weyl 
phase is practically absent. This effect is ascribed to the fact that the 
splittings of both the 
VBM and CBM significantly increase with the increasing Mn concentration. 
This is illustrated in 
Fig. \ref{fig5}, which presents the energies of the 8 highest valence 
states and the 8 lowest 
conduction states at $\Gamma$ as a function of the Sn concentration in 
\pst\ with 0, 2 and 4 
Mn ions in the supercell. Indeed, in the case of \pst\ the spread is the 
smallest, the levels are 
almost degenerate, which is reflected in the wide composition window of 
the Weyl phase. On 
the other hand, in the case of 4 Mn the spread of the levels is 
substantial, their energies are 
well resolved, and the transition is sharp. 
%\onecolumngrid

\begin{figure*}
\includegraphics[width=0.7\linewidth]{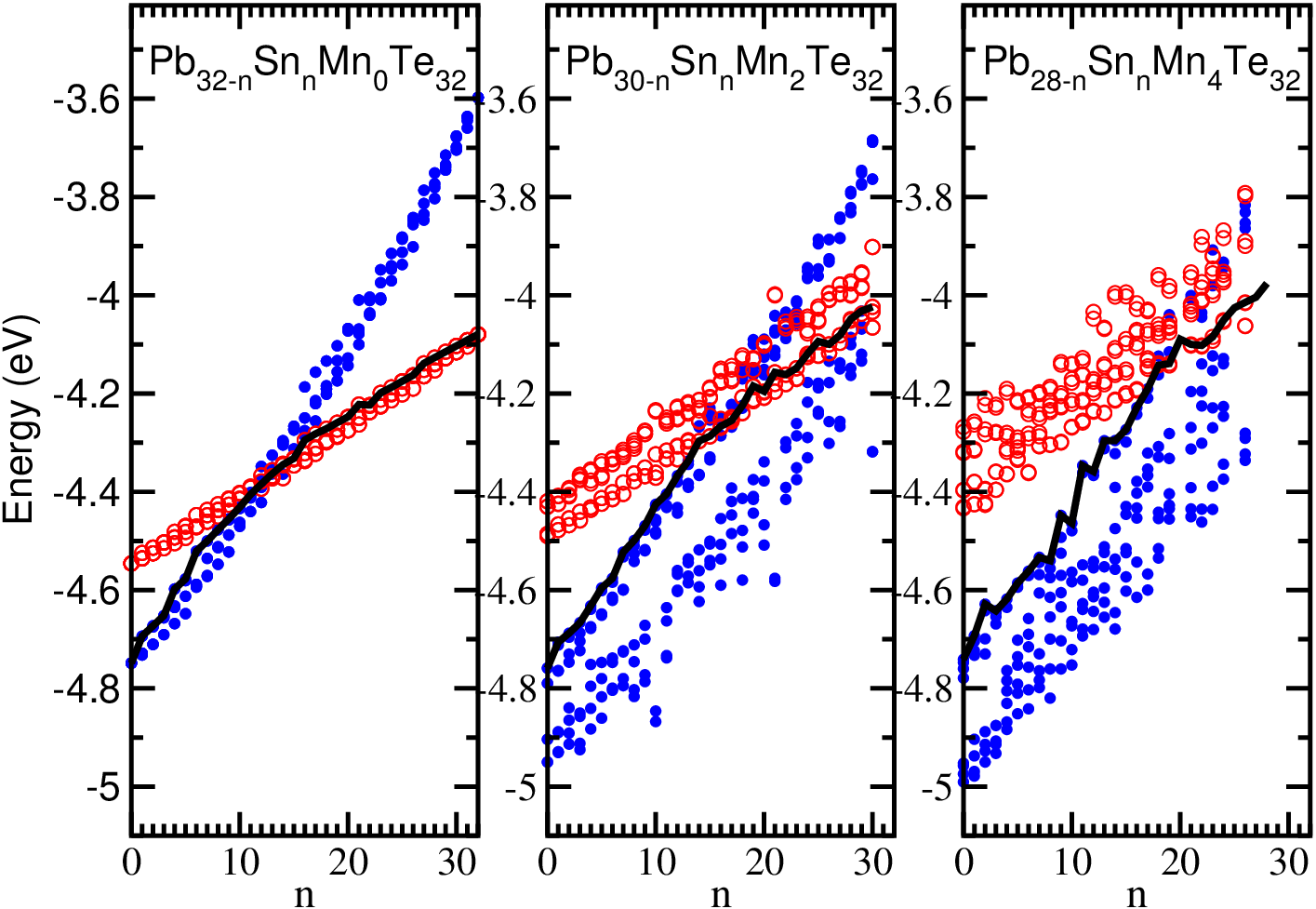}  
\caption{\label{fig5} (color online) Dependencies of 16 energy levels 
nearest to the 
Fermi energy at the $\Gamma$ point of the supercell BZ for different Mn 
concentrations. 
The closed symbols represent the levels with the wave functions dominated 
by the $5p$(Te) orbitals ($L_6^+$ derived states), and the open symbols 
correspond to those 
dominated by the cation $p$ orbitals ($L_6^-$ derived states). Thick 
black lines show the energy position of the VBM.} 
\end{figure*}
%\twocolumngrid

The strong chemical disorder may also be the cause why the Weyl phase is 
usually not 
observed in experiment, with the exception of one work,\cite{ong} in 
which crystals with 
extremely low carrier concentrations were used. In actual alloys, apart 
from the chemical 
disorder, there are also native defects, mostly vacancies, what increases 
the disorder and can 
lead to larger band splittings.

%\onecolumngrid

\begin{figure*}
\includegraphics[width=0.7\linewidth]{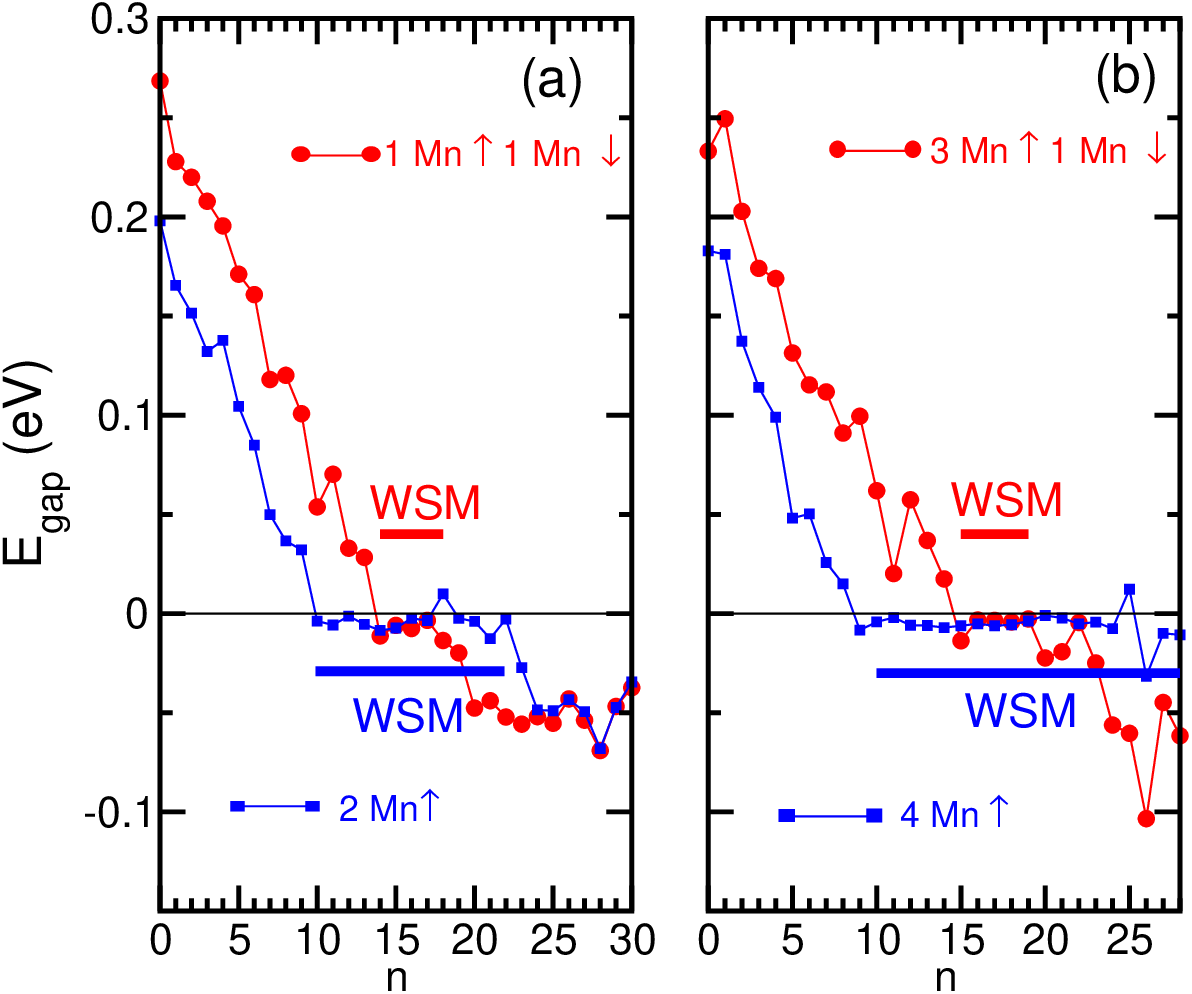}
\caption{\label{fig6} (color online) Dependence of the energy gap on
  the spin polarization for supercells 
containing (a) 2 Mn and (b) 4 Mn atoms. Orientations of Mn spins and spin
configurations are shown by arrows.
Thick horizontal lines denote the regions of the Weyl semimetal phase,
WSM, where the number of Weyl nodes is nonzero. n is the number of Sn
atoms in the supercell.
}
\end{figure*}
%\twocolumngrid

We now turn to the impact of the spin polarization on energy levels in 
\psmt. It was shown in 
detail in Fig. \ref{fig3} for PbTe and SnTe with 2 Mn ions constituting 
BCC lattice. Figure 
\ref{fig6} compare the composition dependence of  the band gap for 
different spin 
configurations for \psmt\ containing two and four manganese atoms in the 
supercells.  In the 
FM configuration, the spin splittings of the VBM and CBM induce a 
considerable reduction of 
the absolute value of $E_{gap}$. As a consequence of spin polarization, 
in the case of alloys 
with 2 Mn ions the Weyl region characterized by $E_{gap}=0$ is 3 times 
wider than that in the 
AFM case. In the case of 4 Mn in the supercells, the effect is even more 
dramatic, since 
$E_{gap}$ vanishes for all composition $x > 0.25$. Indeed, this is in 
sharp contrast with the 
very narrow composition window calculated in the absence of spin 
polarization. These results are in qualitative agreement with those of Ref. \onlinecite{liu},
which also finds that the width of the Weyl region increases with the spin
polarization of magnetic ions.

The above results, together with those for the supercells with 4 Mn ions 
with 3 spins up and 
one spin down, allow for a few quasi-quantitative conclusions. Namely, in the 
case of PbTe, 
incorporation of two Mn ions in the AFM state rises $E_{gap}$ by 0.03 eV, 
while their spin 
polarization lowers $E_{gap}$ by about the same amount. Thus, the presence of 2 
Mn in the FM configuration  
leaves the band gap unchanged, which illustrates well our earlier 
conclusion that the chemical 
and the spin perturbations are equally important. The analogous result 
holds in the case of 4 
Mn ions in supercells. The results for the intermediate case of 4 Mn ions 
with 3 spins up and 
one spin down are fully consistent with this picture. 
\\
In previous works, the trivial-to-TCI phase transition was induced by 
either the change of 
composition of \pst\, or by hydrostatic pressure. The results of Figs. 
\ref{fig4} and \ref{fig6} 
point out to a very interesting possibility of driving the transition by 
applying magnetic field. 
Indeed, in the case of \pst\ with composition x=0.25 with 4 Mn ions, 
$E_{gap}=0.1$ eV in the 
paramagnetic case, i.e., in the AFM configuration, but vanishes in the FM 
configuration. Such a 
scenario is also discussed in Ref. \onlinecite{liu}.

\subsection{\label{results3} The Weyl semimetal phase}
\begin{table}
\caption{\label{tab2} Weyl points $(k_{x,y,z})$ in the BZ of 64-atom 
supercell, the calculated energy gaps $(E_g)$, the corresponding 
topological charges (TC) and energy positions of the Weyl's nodes for   
Pb$_{15}$Sn$_{16}$Mn$_{1}$Te$_{32}$.} 
\begin{ruledtabular}
\begin{tabular}{ccccrr}
  $k_x$ (\AA$^{-1}$) & $k_y$ (\AA$^{-1}$) & $k_z$ 
(\AA$^{-1}$)  &
  $E_g$ (eV) & TC & $E_{val}$ (eV)  \\
  \hline
  \hline
  0.003539 &  0.001528 & 0.000406 & $6.41\times 10^{-
7}$ & -1&-4.2753\\
  0.000002 &  -0.005624 &  0.002274 & $6.75\times 
10^{-7}$ & 1&-4.2780\\
-0.002882  & 0.002512 & -0.003221  &    $3.70\times 
10^{-7}$ & 1&-4.2761\\
0.000131 & -0.005761 & -0.002160   & $1.07\times 
10^{-6}$ & -1&-4.2782\\

\hline
\end{tabular}
\end{ruledtabular}
\end{table}

%% In the transition region the Weyl semimetal phase is expected in \psmt\ 
%% as found previously in \pst. \cite{zunger2} 

A detailed characterization of the Weyl semimetal phase is provided by 
Table \ref{tab2}, where we list the positions of Weyl's nodes in the 
${\bm 
k}$ space, the corresponding energy gaps, topological charges, and the 
energies of the nodes for Pb$_{15}$Sn$_{16}$Mn$_{1}$Te$_{32}$. The 
vanishing $E_{gap}$ makes it impossible to calculate the topological 
indices, as it was discussed in Section II. 

\begin{figure}
\includegraphics[width=\linewidth]{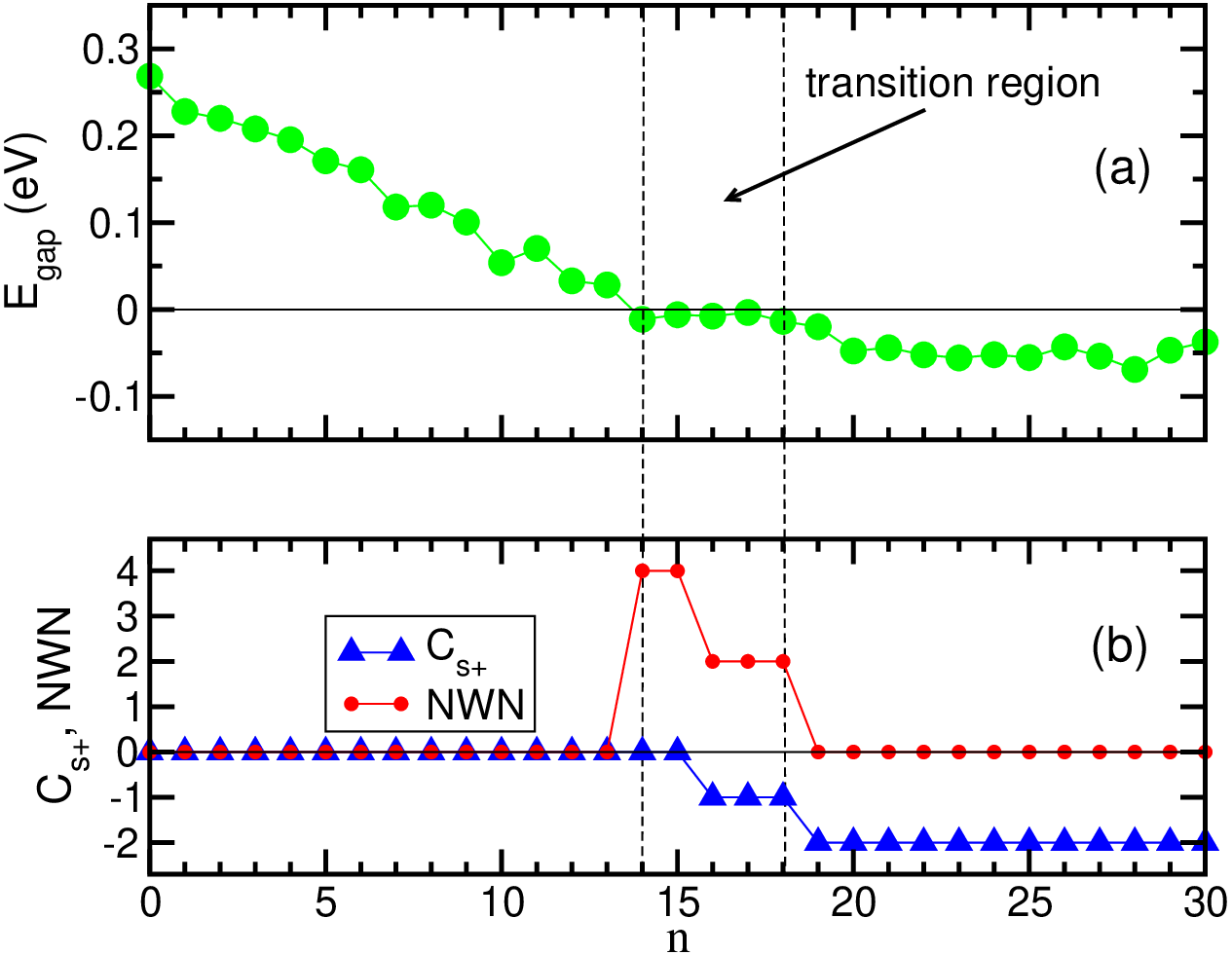}
\caption{\label{fig7}
(color online) The energy gaps (a), the topological indices
$C_{s+}$ and the number of Weyl nodes (NWN) (b) 
for two Mn ions set up antiferromagnetically in  
Pb$_{30-n}$Sn$_{n}$Mn$_2$Te$_{32}$. }   
\end{figure}

As it is shown in Fig.~\ref{fig7}, the analysis of the Weyl's nodes helps 
to 
define more precisely the transition region between the trivial and the 
TCI phase. In the case of 4 Mn ions in the supercell we find only one Sn 
concentration, $n=18$, for which the number of Weyl's nodes is 
nonzero. This confirms the fact that for larger concentrations of Mn the 
Weyl region is very narrow, or is just absent. 

\begin{figure}
\includegraphics[width=\linewidth]{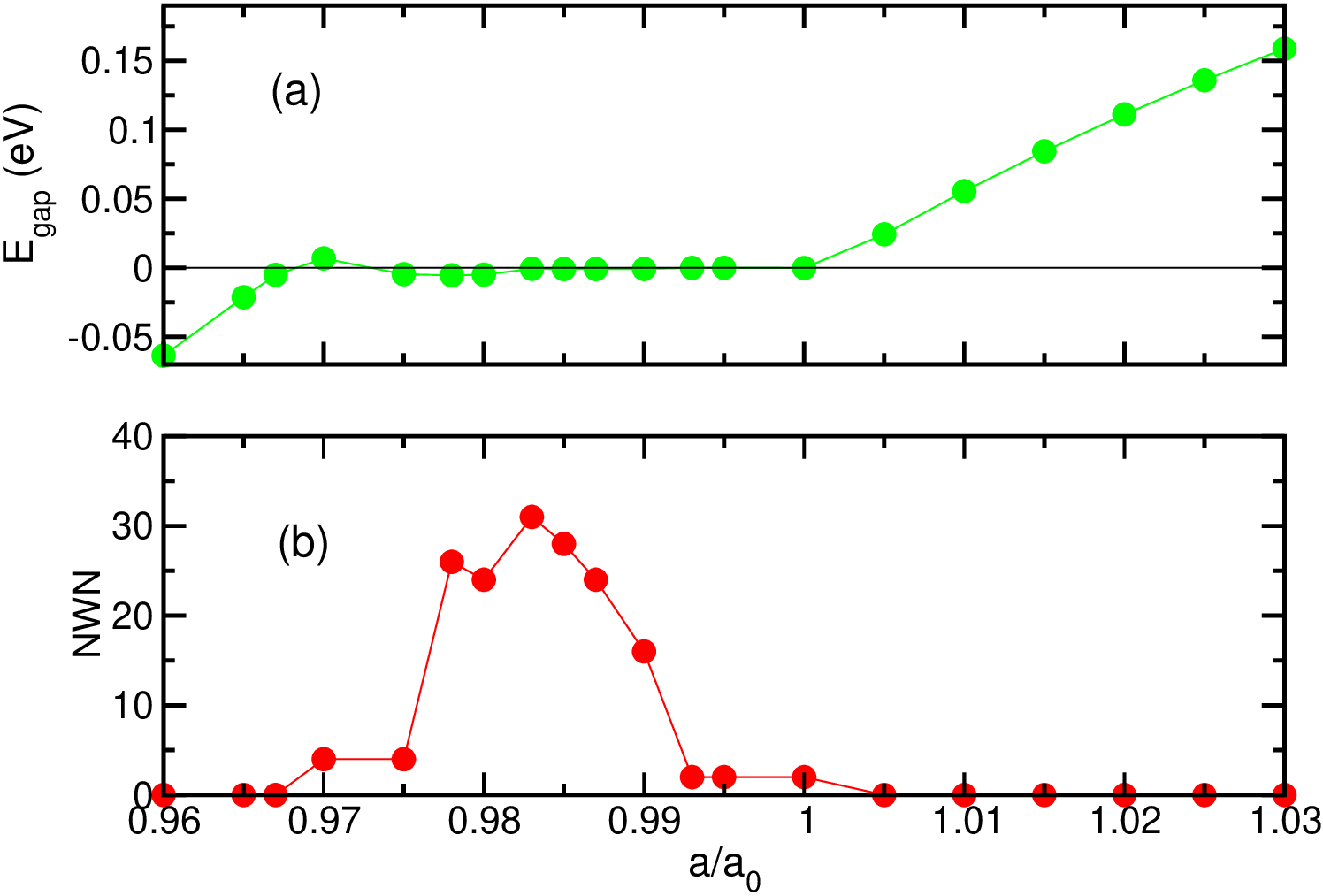}
\caption{\label{fig8} (color online) (a) The energy gap and (b) the number of Weyl nodes (NWN) 
for Pb$_{23}$Sn$_{8}$Mn$_1$Te$_{32}$ as the function of the lattice 
parameter. $a_0$ is the equilibrium lattice parameter of PbTe.}   
\end{figure}

Finally, as additional example, we analyzed the number of Weyl's nodes 
as the function of the lattice parameter for \psmt\ containing one Mn 
ion in the supercell. The results are presented in Fig.~\ref{fig8}. They 
show that the number of Weyl's nodes provides a precise measure of the 
width of the Weyl region.

\section{Conclusions}

With the increasing Sn content in the \pst\ alloy, a transition from the 
topologically trivial to the non-trivial topological crystalline 
insulator (TCI) 
phase takes place. The transition is smeared, because there is a wide 
composition window, in which the alloy has the zero band gap and remains 
in the Weyl semimetal phase. The calculated critical Sn concentration 
corresponding to the onset of the transition of \pst\  to the Weyl phase, 
$x=0.3$, is reasonably close to that experimentally observed. The Weyl 
phase extends from 0.3 to 0.5, and for higher $x$ the alloy assumes the 
TCI 
phase. 

Using ab initio calculations we investigate consequences of alloying 
\pst\  
with Mn. The group-II Mn is chemically different from the Pb and Sn 
group-
IV cations, and thus it introduces a strong chemical perturbation of the 
\pst\ electronic structure. Next, since Mn in the IV-VI compounds is in 
the 
high spin state, perturbation acting on spin variables of band carriers 
is 
present when the macroscopic spin polarization of Mn ions is finite. The 
main conclusions are as follows. 

1. At higher temperatures, the system is paramagnetic with the vanishing 
spin polarization. In this case, the incorporation of Mn ions into \pst\ 
leads 
to the increase of $E_{gap}$ on the PbTe side and its decrease on the 
SnTe 
side, which modifies the composition window of the Weyl phase. At 
sufficiently low temperatures the Mn system in \psmt\ can be spin 
polarized, and the spin splittings of the CBM and the VBM are comparable 
to the band gap, again considerably widening the Weyl area. For example, 
in the presence of $y \approx 0.06$ of Mn, the Weyl area shifts from
$0.3 < x < 0.5$ in \pst\ to $0.5 < x < 0.8$. When the Mn ions are fully 
spin 
polarized, the Weyl area extends from $y=0.5$ to 1.0. 

2. The strong impact of the spin polarization on the energy bands opens 
an 
interesting possibility of inducing a transition from the trivial to the 
Weyl 
phase by magnetic field or by spontaneous magnetization. The effect is 
expected to occur for $x > 0.35$. 

3. \psmt\ alloys can be characterized by topological indices, which are 
based on the concept of the Chern number. If the total spin polarization 
of 
the Mn ions vanishes, the spin Chern number constitutes the appropriate 
topological index. In other  cases, the alloy can 
be 
characterized by two indices, $C_{s+}$ and $C_{s-}$. The dependencies of 
$E_{gap}$ on the Sn content or on the lattice parameter agree very well 
with the corresponding dependencies of our topological indices. Thus, 
they 
constitute a valid characteristics of the system, and in particular they 
reveal 
whether $E_{gap}$ is positive or negative. 

4. In the semimetal Weyl phase, the Weyl's nodes are placed very close 
in 
the ${\bm k}$ space ($\sim $ 0.02 \AA$^{-1}$),  thus the observation of 
the splitting of Dirac cones using ARPES technique is rather not possible at 
present.

\begin{acknowledgments}
This work was partially supported by National Science 
Centre NCN (Poland) projects
UMO-2016/23/B/ST3/03725 (A{\L}), UMO-2017/27/B/ST3/02470 
(A{\L}), and UMO-
2014/15/B/ST3/03833 (TS) as well as by the Foundation 
for Polish Science through the IRA Programme co-financed 
by EU within SG OP (TS).

\end{acknowledgments}

% Authors

\end{document}